\documentclass[a4paper,12pt]{article}

\usepackage[english]{babel}
\usepackage{authblk}
\usepackage{dsfont}
\usepackage{amsfonts}
\usepackage{mathrsfs}
\usepackage{amssymb}        
\usepackage{amsmath}
\usepackage{graphicx,caption}
\usepackage{float}
\usepackage[a4paper]{geometry} 
\usepackage{verbatim}
\usepackage{pstricks}
\usepackage{amsthm}
\usepackage{mathrsfs}

\newcommand{\R}{\mathbb{R}}

\newcommand{\Z}{\mathbb{Z}}

\newcommand{\dd}{\mathrm{d}}

\newcommand{\ed}{\mathrm{e}}

\begin{document}

% title
\title{Glassy dynamics in strongly anharmonic chains of oscillators}

\author[1]{Wojciech De Roeck}
\affil[1]{
Instituut Theoretische Fysica, 
KU Leuven, 
3001 Leuven, Belgium
}

\author[2]{Fran\c cois Huveneers}
\affil[2]{
Ceremade,
UMR-CNRS 7534, 
Universit\'e Paris Dauphine, 
PSL Research University, 
%Place du Mar\'echal de Lattre de Tassigny, 
75775 Paris cedex 16, 
France
}

\date{\today}
\maketitle 

\begin{abstract}
\noindent
We review the mechanism for transport in strongly anharmonic chains of oscillators near the atomic limit where all oscillators are decoupled.
In this regime, the motion of most oscillators remains close to integrable, i.e.\@ quasi-periodic, on very long time scales, 
while a few chaotic spots move very slowly and redistribute the energy across the system.  
The material acquires several characteristic properties of dynamical glasses: 
intermittency, jamming and a drastic reduction of the mobility as a function of the thermodynamical parameters. 
We consider both classical and quantum systems, though with more emphasis on the former, and we discuss also the connections with quenched disordered systems, 
which display a similar physics to a large extent. 
\end{abstract}

\section{Introduction}
In an isotropic system with a single locally conserved quantity, the energy, 
Fourier's law asserts that the heat flux $J$ is proportional to the temperature gradient: 
\begin{equation}\label{eq: Fourier law}
	J \; = \;  \kappa (\beta) \nabla \beta \, 
\end{equation}
where $\beta = (k_B T)^{-1}$ denotes the inverse temperature and where the scalar quantity $\kappa (\beta)$ is the thermal conductivity.\footnote{
It is more conventional, but slightly less convenient here, to define the thermal conductivity through the relation $J = - \tilde \kappa(T) \nabla T$
with $k_B \tilde \kappa (T) = \beta^2 \kappa (\beta)$.
}
This is a macroscopic law that becomes valid at the scales where the system can locally be described by a Gibbs state at inverse temperature $\beta$. 
Based on general thermodynamical arguments, one concludes that $\kappa (\beta)\ge 0$, 
since the entropy production rate $\gamma_{s}$ is given by  
\begin{equation*}%\label{eq: entropy porduction rate}
	\gamma_s \; = \; \nabla \beta \cdot J \; = \; \nabla \beta \cdot \kappa(\beta) \nabla \beta \; \ge \; 0. 
\end{equation*}
This is however the only general property of $\kappa (\beta)$ that can be derived from entropic considerations.  
As clearly realized by Green \cite{green_1954} and Kubo \cite{kubo_1957,kubo_1966}, 
the value of the conductivity does depend on the microscopic dynamics through some dynamical correlations. 
It may be explicitly expressed as 
\begin{equation}\label{eq: Green Kubo}
	\kappa (\beta) \; = \; 2\int_0^\infty \dd t \int \dd x \, \langle J(x,t) \cdot J (0,0) \rangle_\beta
\end{equation}
where $\langle \cdot \rangle_\beta$ denotes the equilibrium state at inverse temperature $\beta$, 
and where $J$ represents now the current on the microscopic scale.
The current-current correlator featuring in \eqref{eq: Green Kubo} 
may display a wealth of interesting phenomena that cannot be diagnosed by inspection of the equilibrium states alone. 
  
In this review, we consider systems where diffusive transport on long space-time scales is expected, and we will mostly take this for granted. 
Instead, we focus on intermediate scales
and we analyze the finite-time effects that determine the value of the conductivity, as predicted by \eqref{eq: Green Kubo}. 
The systems that we will analyze are chains of classical or quantum oscillators with a Hamiltonian of the form $H = H_0 + g H_1$. 
At $g = 0$, all oscillators are uncoupled and no conduction is possible. 
Moreover, we assume that the dynamics of each individual oscillator is integrable and anharmonic. 
For a classical oscillator, anharmonicity means that the frequency genuinly varies with energy, 
and for a quantum oscillator, it means that the level spacing is non-constant, see below for concrete examples.
When $g > 0$, the Hamiltonian $H_1$ provides a coupling between the oscillators, 
but the anharmonicity prevents efficient transport of energy due to frequency mismatch, except in rare places called resonances where there is accidentally no mismatch.
These resonances play a crucial role in the energy transfer, 
since they may host chaotic spots that will eventually break the quasi-periodic behavior and ensure the validity of Fourier's law \eqref{eq: Fourier law} in the long run.     

The analysis discussed below yields a completely different picture for transport 
than the one involving weakly interacting phonons, that prevails near the harmonic (ballistic) limit 
\cite{lefevere_schenkel_2006,aoki_lukkarinen_spohn_2006,lukkarinen_2016}. 
Indeed, close to $g=0$, our systems exhibit several characteristics of dynamical glasses: 
strongly intermittent dynamics with only a very low density of active spots, 
constrained motion of the chaotic spots due to the random energy landscape, 
and drastic fall-off of their mobility as one approaches the limit $g=0$
(which translates in most systems of interest into some limit for the temperature or some other thermodynamical parameter). 
This behavior was unveiled
in~\cite{oganesyan_pal_huse_2009,basko_2011,de_roeck_huveneers_2014,de_roeck_huveneers_2015,de_roeck_huveneers_pspde_2015,
de_roeck_huveneers_muller_schiulaz_2016,huveneers_adP_2017} 
and was recently checked through a detailed numerical analysis in~\cite{danieli_et_al_2018,mithun_et_al_2019}. 
Hence, the poor transport properties of systems close to the atomic limit is reasonably understood. Yet, 
the theory still lacks testable quantitative predictions for the time scales on which transport eventually sets in, 
despite the detailed analysis in \cite{basko_2011} 
for the disordered discrete non-linear Schr\"odinger (DDNLS) chain, and the (related) one in \cite{huveneers_adP_2017} for  the translation invariant rotor chain. 

That frequency mismatch inhibits the transfer of energy is by itself a well-known physical phenomenon. 
In classical few-body Hamiltonian systems, 
the KAM theorem shows that, at small coupling, a large proportion of the phase space is occupied by invariant tori where the motion is quasi-periodic, 
see e.g.~\cite{poschel_2001},
while Nekhoroshev's estimates guarantee the conservation up to stretched exponentially long times in the perturbative parameter of the uncoupled actions, 
see e.g.~\cite{poschel_1993}. 
In quantum mechanics, disorder may strongly suppress quantum tunneling and lead to the localization in space of the wave packet, as first discovered by Anderson
\cite{anderson_1958}. 
These phenomena are rather fragile though, and they are not generally expected to survive in the presence of dissipative effects, e.g.\ coupling to an environment.  
Therefore, if we consider extended systems at positive values of the thermodynamical parameters, 
it is a priori all but clear whether and how they those phenomena can persist.  
A careful investigation started in the mid 2000s for disordered quantum systems, opening the field of many-body localization (MBL)
\cite{fleishman_anderson_1980,gornyi_mirlin_polyakov_2005,basko_aleiner_altshuler_2006}. 
Remarkably, the localization in some systems was found to be so robust that it persists in the thermodynamic limit
\cite{oganesyan_huse_2007,imbrie_jsp_2016,imbrie_prl_2016}.
Such systems remain however rather scarce and, 
as we stressed above, we will deal in this review with a much broader class of systems where any form of integrability is eventually washed out by dissipative effects. 
  
The rest of this review is organized as follows. 
For convenience, we restrict ourselves to one-dimensional systems, but none of the considerations developed here depends on the dimension. 
In particular, poor transport is not due to the possible presence of bottlenecks in the chain 
\cite{agarwal_gopalakrishnan_knap_mueller_demler_2015,gopalakrishnan_agarwal_demler_huse_knap_2016,agarwal_et_al_2017}, 
though these effects may be present and constitute an additional slowing down factor. 
The three next sections deal with classical systems. 
In Section~\ref{sec: rotor chain}, we study the rotor chain, or classical Josephson junction chain, as a generic example. 
Generalizations, together with some related questions, are discussed in Section~\ref{sec: more on classical}. 
In Section~\ref{sec: disordered systems}, we make the connection with quenched disordered systems. 
We discuss quantum systems in Section~\ref{sec: quantum systems}.

\section{The rotor chain}\label{sec: rotor chain}
Let us consider the classical Hamiltonian 
\begin{equation}\label{eq: rotors}
	H (\theta, \omega) \; = \; \frac{m}{2}\sum_{x = 1}^L \omega_x^2 + g \sum_{x=1}^{L} \left( 1 - \cos (\theta_{x+1} - \theta_x)  \right)
\end{equation}
where $L$ is the length of the system, 
where $\theta = (\theta_x)_{1\le x \le L}$ with $\theta_x \in \R / 2\pi \Z$ are angles
and where $\omega = (\omega_x)_{1\le x \le L}$ with $\omega_x \in \R$ are angular velocities.
For convenience, we assume periodic boundary conditions $\theta_{L+1} = \theta_1$, and we take units of mass such that $m=1$. 
This system has two locally conserved quantities: the energy $H$ and the angular velocity $\Omega = \sum_{x=1}^L \omega_x$. 
Let us consider the system at equilibrium at inverse temperature $\beta$ and zero average velocity, 
as can always be assumed without loss of generality by performing a Galilean boost.  
The variables $\theta$ and $\omega$ are conjugated and the dynamics is given by Hamilton's equations: 
\begin{equation}\label{eq: hamilton rotors}
	\dot \theta_x \; = \; \omega_x, 
	\qquad 
	\dot \omega_x \; = \; g \left( \sin (\theta_{x+1} - \theta_x) - \sin (\theta_x - \theta_{x-1}) \right) 
\end{equation}

In the limit $g = 0$, the system reduces to independent oscillators that are trivially integrable: 
the angular velocity of each of them is conserved, $\omega_x (t) = \omega_x (0)$, and the angles evolve as $\theta_x (t) = \theta_x (0) + \omega_x(0) t$. 
Let us next consider $g > 0$ small enough such that $\beta g \ll 1$. 
Though the angular velocities now evolve with time, their statistics remains unchanged in the Gibbs state: 
$\omega_x$ are independent and identically distributed (iid) Gaussian random variables with mean 0 and variance $\beta^{-1}$.
% Natural unit of time in this problem: $(m \beta)^{1/2} = 1$. This is the time for one typical rotor to perform one full rotation
Moreover, in first order approximation in $g$, the evolution of $\omega_x$ is given by 
\begin{equation}\label{eq: first order correction}
	\omega_x (t)
	\; \simeq \; 
	\omega_x (0) + g \int_0^t \dd s \, \left(  \sin (\Delta\omega_{x+1}(0) s) - \sin (\Delta\omega_{x}(0) s) \right) 
\end{equation}
with $\Delta \omega_x = \omega_{x} - \omega_{x-1}$.
The integral featuring in~\eqref{eq: first order correction} is oscillatory and averages to $0$, 
so that it yields at all times a correction for the angles with typical value of order $\beta g$.   
It is however crucial to realize the possibility of atypical behavior: 
First order resonances occur when one of the differences $\Delta\omega_{x+1}(0)$ or $\Delta\omega_{x}(0)$ is atypically small, 
i.e.\@ of order $g^{1/2}$ or smaller.
In this specific system, due to the presence of a second conserved quantity, two coupled oscillators do still form together an integrable system 
(that can be mapped to a pendulum) and the above resonances cannot by themselves constitute a mechanism to transport the energy. 
However, the dynamics of three consecutive oscillators with their angular velocity in a band of width $g^{1/2}$ is chaotic. 
Such a configuration is called a chaotic triple, and it occurs with a probability $\beta g$. 
See \cite{chirikov_1979} for a deeper discussion on the origin of chaos in Hamiltonian systems, 
and \cite{basko_2011,huveneers_adP_2017} for specific applications relevant here.  

\begin{figure}[t]
    	\centering
   		\includegraphics[draft=false,height = 10cm,width = 15cm]{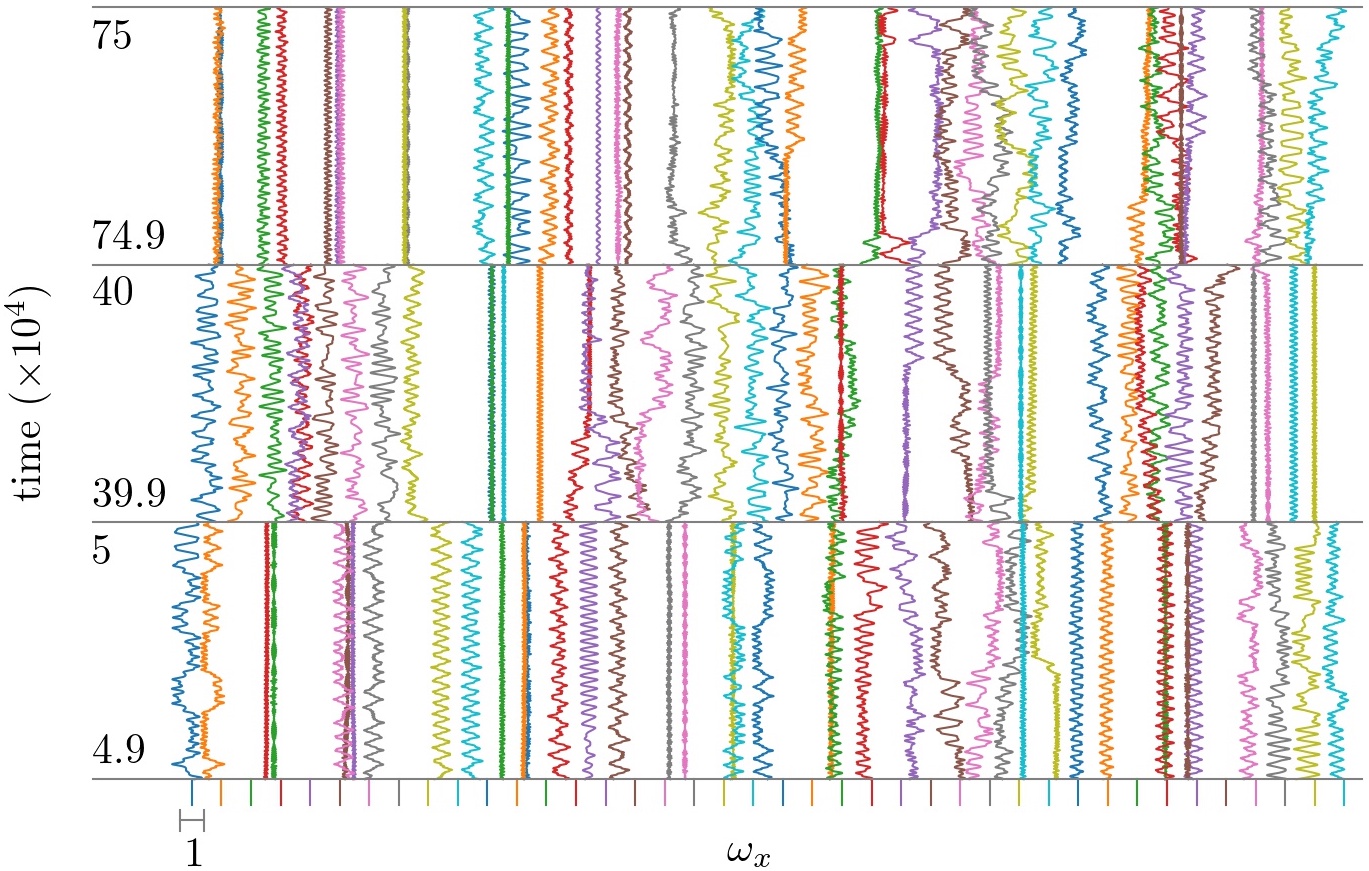}
		\captionsetup{width=.9\linewidth}
    	\caption{
		Three snapshots of the evolution of the angular velocities $\omega_x$ for $x=1$ to $x=40$ 
		for the Hamiltonian \eqref{eq: rotors} at thermal equilibrium, 
		with $L = 64$, $m=1$, $\beta = 1$ and $g = 0.07$.
		Vertical tick marks below correspond to the values $\omega_x = 0$ for each oscillator. 
		} 
    	\label{fig: individual trajectory}
\end{figure}

\begin{figure}[t]
    	\centering
   		\includegraphics[draft=false,height = 5.5cm,width = 8cm]{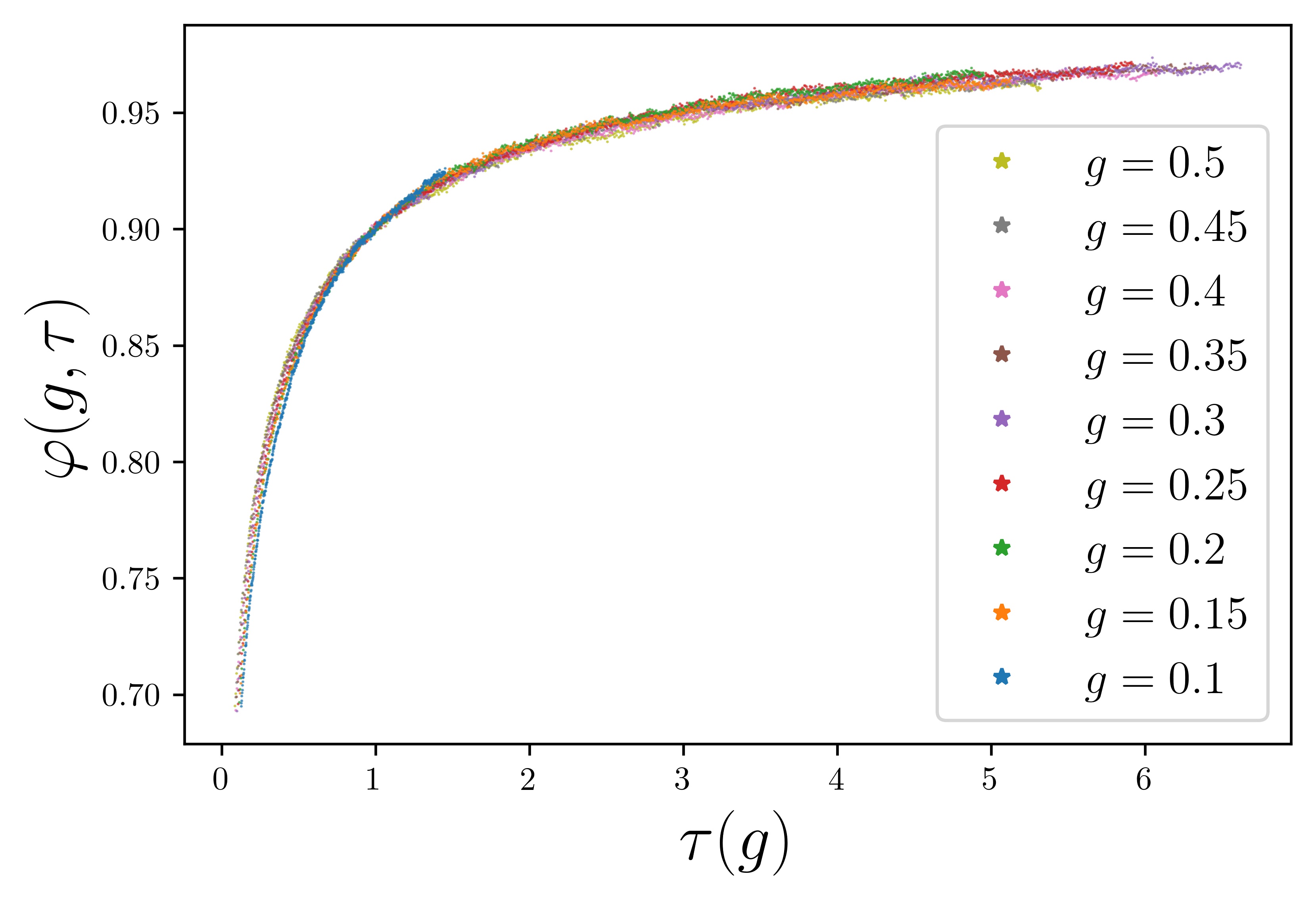}
		\hspace{0.5cm}
		\includegraphics[draft=false,height = 5.5cm,width = 8cm]{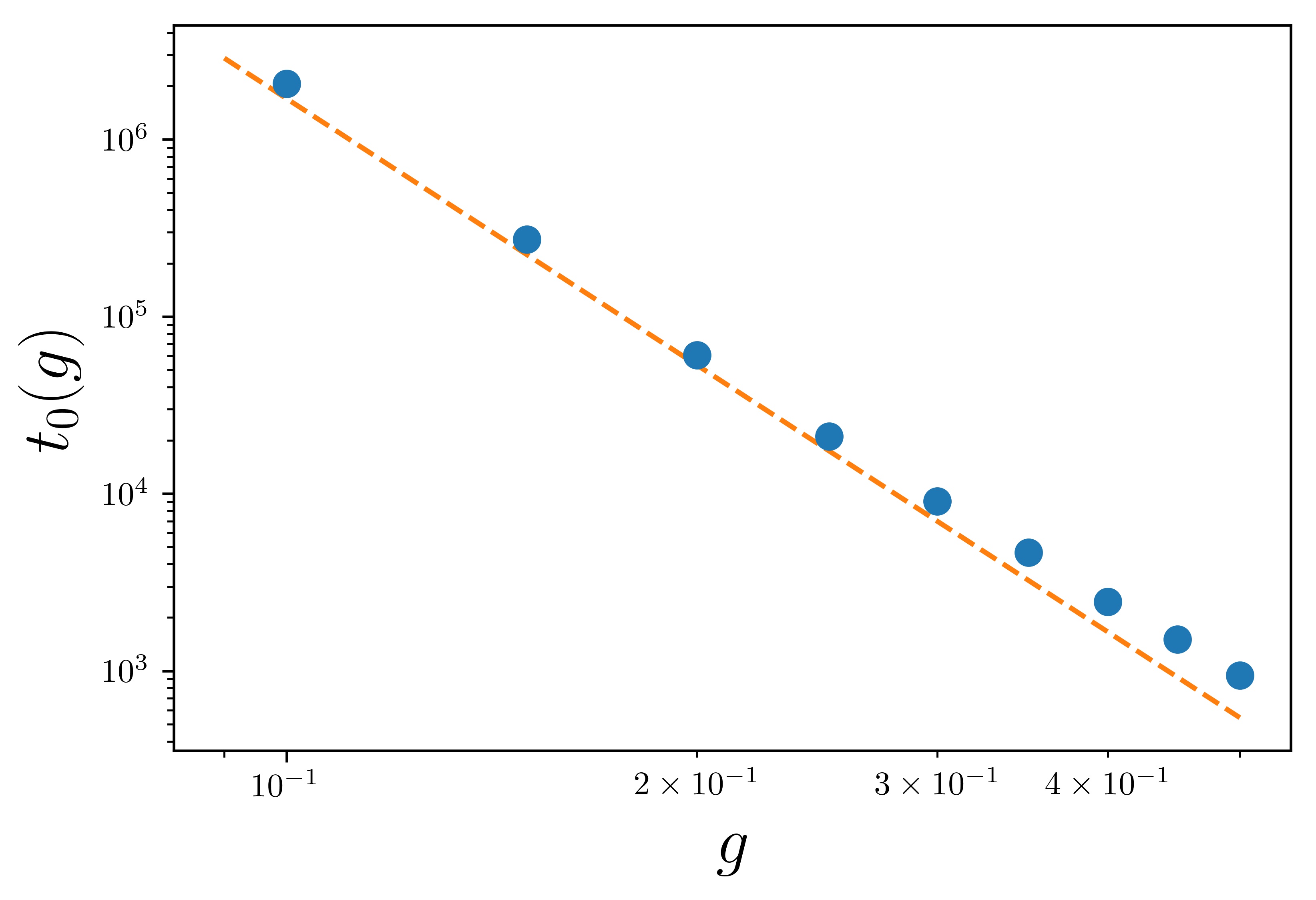}
		\captionsetup{width=.9\linewidth}
    	\caption{
		Panel (a). Time evolution of $\varphi$ as a function of the rescaled time $\tau (g)$ for various values of $g$ (see main text), 
		for $L = 64$, $\beta = 1$ and $m = 1$.
		Panel (b). Relaxation time $t_0 (g)$ for the same system (see main text). The dotted line with slope $-5$ serves as a guide for the eye. 
		Average over 8000 initial configurations at least for each value of $g$.
				} 
    	\label{fig: thermalization parameter}
\end{figure}

From this crude first order analysis, one concludes already that the dynamics displays an intermittent behavior in the regime $\beta g \ll 1$
(at least as long as first order corrections dominate, but we will see that the conclusion extends to much longer times): 
Most oscillators display a quasi-periodic behavior while rare resonant triples act as heat baths and redistribute the energy into the chain.    
Intermittency is illustrated on Fig.~\ref{fig: individual trajectory}, where three snapshots of the dynamics at different times are shown.\footnote{ 
In our simulations,
the system is first put in thermal equilibrium by acting on each coupled oscillators with a Langevin heat bath at inverse temperature $\beta$ over a time $t_{th}=1000$, 
and the isolated dynamics is then simulated with a St\"ormer-Verlet scheme 
with an elementary time step $\Delta t = 0.001$ for Fig.~\ref{fig: individual trajectory}
and $\Delta t = 0.01$ for Fig.~\ref{fig: thermalization parameter}. 
}
See also Fig.~4 in \cite{mithun_et_al_2019}. 

Let us next analyze how the chaotic triples may induce transport. 
Since these spots occur with a probability $\beta g$, a contribution of order $(\beta g)^2$ to the conductivity could be a first naive guess, 
but it turns out to be much smaller. 
To see it, following \cite{basko_2011}, we contemplate two mechanisms for the transfer of energy and, incidentally, 
we notice already that each of them will play a distinguished role in different quantum systems, see Section~\ref{sec: quantum systems}.
First, fixed spots may exchange energy among themselves. 
However, since they are far appart and since the material in between is assumed to contain no chaotic region, 
this will require higher order processes, involving typically $(\beta g)^{-1}$ oscillators. 
Second, spots may redistribute the energy in the oscillators in their vicinity and start moving across the chain, 
through the formation and the destruction of resonances.
However, the process of transferring energy between the spot and a site at a relative angular velocity $\Delta\omega$ compared to the average angular velocity of the spot
is reduced by a factor $\ed^{-\Delta\omega / g^{1/2}} \sim \ed^{-(\beta g)^{-1/2}}$ typically. 
This reduction also ``explains" the validity of Nekhoroshev's estimates. 

We came thus to the conclusion that there is no transport due to first order effects in $\beta g$ (including first order resonant spots). 
It is possible to undertake a systematic order-by-order analysis \cite{basko_2011,Huveneers_2012} of higher order analogues of \eqref{eq: first order correction}. 
%
% Such a procedure involves , which 
%are dominant in the absence of higher order resonances that involve typically several sites and host chaotic spots.}
The conclusion remains always the same: there is no transport at any given order, 
but mechanisms are identified that will lead eventually to the transfer of energy on longer time scales. 
By an optimization procedure first carried over in \cite{basko_2011} for the DDNLS chain, one comes to a prediction for the rate $\gamma$ at which energy is transferred:
\begin{equation}\label{eq: basko rate}
	\gamma \; \sim \; g^{1/2} \, \ed^{ - c (\log (\beta g))^2}, \qquad c > 0 \, ,
\end{equation}
see also~\cite{huveneers_adP_2017}. 
As it turns out, the most efficient channel for transport is the second one described above, 
i.e.\@ chaotic spots {move around with a rate} proportional to $\gamma$ given by \eqref{eq: basko rate} and they modify the energy landscape in their vicinity. 
This mechanism, as well as the very small value of $\gamma$, properly reveal the glassy dynamics of the rotor chain in the regime $\beta g \ll 1$. 

It would be an overkill to take the analytical expression \eqref{eq: basko rate} literally, 
since its derivation involves several approximations and since it would be exceedingly hard to validate it numerically, let alone experimentally. 
It has nevertheless the virtue to highlighting two robust pieces of information. 
First, at finite volume, the rate $\gamma$ would be a stretched exponential in the inverse coupling parameter, as follows from Nekhoroshev's estimates~\cite{poschel_1993}.
Rigorous upper bounds at finite volume yield a power in the stretched exponential decaying as $1/L$ as $L \to \infty$,
and the expression \eqref{eq: basko rate} confirms that this power indeed vanishes in the thermodynamic limit.  
We notice in passing the rigorous result in \cite{benettin_froehlich_giorgilli_1988}: 
An upper bound for the rate $\gamma$ for an initially localized wave packet in an infinite lattice (i.e.\ at zero density) is derived and it basically coincides with \eqref{eq: basko rate}. 

Second, \eqref{eq: basko rate} shows that the rate $\gamma$ decays faster than any polynomial in $\beta g$ in the limit $\beta g \to 0$.
This leads to the claim that, in this limit, the conductivity is non-perturbative or non-analytic {at origin} as a function of $\beta g$, 
see \cite{de_roeck_huveneers_2014,de_roeck_huveneers_2015}. 
This fact is supported by several mathematical results, discussed below, and should be kept in mind when interpreting numerical results. 
As an example, let us report on numerical simulations of a simple quantity aimed to capture the relaxation time in the system: 
$$
	\varphi (g,\beta,t) \; = \; \lim_{L\to \infty} \frac{\beta}{2L} \sum_{x=1}^L \langle (\omega_x(t) - \omega_x(0))^2\rangle_\beta. 
$$ 
Clearly, $\varphi (g,\beta,0) = 0$ and we expect that $\varphi (g,\beta,t) \to 1$ as $t \to \infty$.
Thanks to the scaling relation $\varphi(g,\beta,t) = \varphi (g/\alpha^2 , \beta \alpha^2 ,\alpha t)$ valid for any $\alpha > 0$,\footnote{
This follows from the scaling relations 
$\theta (t,g;\theta^o,\omega^o) = \theta(\alpha t,g/\alpha^2;\theta^o,\omega^o/\alpha)$
and 
$\omega (t,g;\theta^o,\omega^o) = \alpha \omega(\alpha t,g/\alpha^2;\theta^o,\omega^o/\alpha)$, 
with $(\theta^o,\omega^o)$ the initial conditions and $\alpha > 0$,
that are derived from \eqref{eq: hamilton rotors}. 
}
it suffices to analyze $\varphi(g,\beta,t)$ at a given value of $\beta$ and vary $g$ (and we further drop $\beta$ from the notations). 
In Fig.~\ref{fig: thermalization parameter}(a), we show the the time evolution of $\varphi$ for different values of $g$
as a function of the rescaled time $\tau(g) = t/t_0(g)$ with $t_0(g) = \min\{ t: \varphi(g,t)=0.9\}$. 
Under this rescaling, we observe that the data fit very well with each others, 
despite some very small but neat deviation for the smallest value $g = 0.1$, 
and we conclude that $t_0(g)$ can serve as a meaningful indicator of the relaxation time of the dynamics.\footnote{
The data strongly suggest that $\varphi$ grows monotonically as a function of $t$. 
To determine $t_0$, we have averaged the values of $\varphi$ in small intervals of time so as to get smooth monotonic curves. 
}
The dependence of $t_0$ on $g$ is shown on Fig.~\ref{fig: thermalization parameter}(b) and a fit with a power law yields $t_0 (g) \sim g^{-5}$.\footnote{
We did not find any dependence on the system size $L$ and this was also not expected: 
Since the angular velocity is conserved, we expect that $1 - \varphi(g,t) \sim c(g) t^{-1/2}$ as $L \to \infty$.
}
This is in line with recent numerical results in \cite{mithun_et_al_2019} 
where a similar parameter measuring the ergodization time is found to scale as $g^{-6.5}$. 
One can interpret these observed high powers as a confirmation of the drastic increase of time scales in \eqref{eq: basko rate} as $g \to 0$. 
If taken literally, these power laws would actually contradict \eqref{eq: basko rate} and the non-perturbative nature of transport.
However, since the bounds on the perturbative contributions to transport are actually mathematically rigorous (see below), we think that what is at play here is the numerical difficulty of distinguishing a high power from a superpolynomial function. Stated differently, it is hard to know whether the proper asymptotic regime is reached.

Finally, we would like to stress that the non-analytical behavior of the conductivity as $\beta g$ approaches $0$
can be justified mathematically to a large extent, though not entirely.
To get a tractable mathematical problem, it is convenient to introduce some noise in the system, 
following a strategy developed in \cite{basile_bernardin_olla_2006} in the context of heat conduction. 
The presence of noise is unavoidable in any real or even numerical experiment and, 
since the absence of transport is usually associated to integrable effects, we expect the noise to enhance the dissipation of the energy.
The upper bound below on the conductivity for the noisy system is thus expected to hold for the isolated system too.

We introduce a noise that conserves the total energy (but incidentally not the angular velocity), as it simplifies the theoretical analysis of the conductivity. 
Practically, the noise flips independently the velocity of the oscillators at a rate $\lambda$.
Formally, the full dynamics is now described by the stochastic generator $\mathcal L = \mathcal A + \lambda \mathcal S$, 
where $\mathcal A = \omega\cdot \nabla_\theta - \nabla_\theta H \cdot \nabla_\omega$ is the usual Liouville operator and
\begin{equation}\label{eq: velocity flip}
	\mathcal S f(\theta,\omega) \; = \; \sum_{x=1}^L \big(  f(\theta,\omega_1, \dots , - \omega_x, \dots , \omega_L) - f (\theta, \omega) \big).
\end{equation}
Let us assume that $\beta g \ll 1$ as before, and let us set
\begin{equation}\label{eq: strength noise}
	\lambda \; = \; g^{1/2} (\beta g)^r \qquad \text{for} \qquad r \gg 1 
\end{equation}
i.e.\@ $\lambda$ is a high power of $\beta g$. 
The mathematical result shown in \cite{de_roeck_huveneers_2015} claims that, 
for fixed $\beta>0$ and for any $r>0$ there exists a constant $C_r(\beta) < + \infty$ so that
\begin{equation}\label{eq: math conductivity}
	\kappa (g,\beta,\lambda) \le C_r(\beta) g^2 \lambda.
\end{equation}
{The interpretation of this bound is clear: There is no mechanism for transport that is stronger than the added noise with strength $\lambda$.}\\
Back to the system at $\lambda = 0$. It is convenient to deal with the reduced dimensionless conductivity $\overline \kappa (g,\beta) = g^{-1/2}\beta^2 \kappa (g,\beta)$
since scaling relations used above yield here that $\overline \kappa$ depends only on $g,\beta$ through the product $\beta g$.
Loosing strict mathematical rigor, we infer from \eqref{eq: math conductivity} that 
\begin{equation}\label{eq: non math conductivity}
	\overline \kappa (\beta g) \; \le \; C_r (\beta g)^r \qquad \text{for all } r > 0. 
\end{equation}
{The only way that the latter bound would fail is if there was actually a transport mechanism that was destroyed by the added noise of strength $\lambda$, something that we find very hard to imagine in chaotic diffusive systems\footnote{This would, of course,  make sense in integrable ballistic systems, but that is not the focus here.}.}

\section{Translation invariant classical systems}\label{sec: more on classical}
The phenomenology discussed above for the case of a chain of rotors applies, or is expected to do, 
to a large class of translation invariant classical chains of oscillators.  
In \cite{de_roeck_huveneers_2015}, the discrete non-linear Schr\"odinger (DNLS) chain
\begin{equation}\label{eq: dnls}
	H (\psi,\overline\psi) 
	\; = \;
	U \sum_{x = 1}^L |\psi_x|^4 + g \sum_{x = 1}^L |\psi_{x+1} - \psi_x|^2 \, ,
\end{equation}
with the dynamics $i \dot\psi = \nabla_{\overline{\psi}} H$,
is considered in the regime $g/U \ll 1$ or analogously for large value of the action density $\alpha = \sum_{x=1}^L |\psi_x|^2 / L$. 
Results comparable to \eqref{eq: math conductivity} and \eqref{eq: non math conductivity} are found. 
Since both the rotor chain \eqref{eq: rotors} and the DNLS chain \eqref{eq: dnls} do have two conserved quantities, 
it is worth stressing that it did not play any role in the above discussion. 
% and that, to stress it, the second conservation law is explicitly broken in \cite{de_roeck_huveneers_2015}. 
In particular, the glassy behavior that we discuss here should be distinguished from the one that may arise due to the frustration of two conserved quantities, 
even though they have a common dynamical origin. 
In the DNLS chain \eqref{eq: dnls}, if the energy density $\varepsilon = H/L$ exceeds a certain threshold for a given value of the action density $\alpha$, 
the system enters in an effective negative temperature phase and forms a macroscopic condensate, 
or dynamical breather \cite{rumpf_2004,iubini_et_al_2013,chatterjee_2017,mithun_kati_danieli_flach_2018,huveneers_theil_2019}. 
A breather turns out to be stable on extremely long time scales, due to a very strong frequency mismatch, and leads to a very slow approach towards equilibrium. 
In contrast to the phenomenology considered in this review, the formation of such breathers in the effective negative temperature phase is driven by entropic forces. 
It does not need to entail slow evolution for all observables of interest, 
and in particular the thermal conductivity in this phase is expected to behave just as the conductivity in the usual infinite temperature phase 
(i.e.\@ a Gibbs state with no constraint on the energy), at least in $d > 1$ where no bottleneck effects are present. 

More generally, we may consider oscillator chains of the form 
\begin{equation}\label{eq: usual chain}
	H (p,q) \; = \; \sum_{x=1}^L \left( \frac{p_x^2}{2m} + U(q_x) + V (q_{x+1} - q_x) \right)
\end{equation}
where the momenta $p$ and the positions $q$ are conjugated variables, and e.g.\@ $q_{L+1} = q_L$ (free boundary conditions). 
We expect bad transport properties at high temperature if the on-site potential $U$ grows faster at infinity than the coupling potential $V$:
\begin{equation}\label{eq: bad transport condition}
	U(q) \;\sim\; q^a
	\quad\text{and}\quad
	V(q) \;\sim\; q^b
	\quad \text{as}\quad 
	q \to \infty
	\qquad \text{with}\qquad
	a > b \ge 2
\end{equation}
where the condition $a > 2$ guarantees that the on-site frequency grows with the energy. 
To our knowledge, 
transport in the chain \eqref{eq: usual chain} under the conditions \eqref{eq: bad transport condition} has not been properly investigated so far, 
but several indirect studies are worth pointing out.

One way to understand local dissipative properties of extended systems is to take a small chain and connect it at the boundary to perfect heat baths, 
modeled e.g.\@ by Langevin stochastic thermostats. 
In \cite{hairer_mattingly_2009}, the chain \eqref{eq: usual chain} is considered for polynomial potentials with $a > b=2$ in \eqref{eq: bad transport condition}. 
An effective model for the dissipation of the energy into the baths is derived, 
assuming that there is a single dynamical breather, i.e.\@ a single oscillator at atypically high energy, initially in the middle of the chain: 
\begin{equation}\label{eq: effective dissipation h-m}
	\frac{d H}{dt} \sim  - H^{- f(L)}
	\qquad \text{with} \qquad 
	f(L) \; = \; \left( 1 - \frac{2}{a} \right) (L - 1) - 1 \, .
\end{equation}
% and for $L$ odd (an analogous expression for $L$ even holds obviously as well).
Numerical tests show a very good agreement between the prediction from \eqref{eq: effective dissipation h-m} and the true behavior of the system.
It should be noticed that, despite the appearance of the length $L$ in this expression, 
it is only valid in the regime where $L$ is fixed first and the energy of the breather is then taken to be very large.
A similar analysis is performed for the rotor chain \eqref{eq: rotors} in \cite{cuneo_eckmann_wayne_2017}
coupled on one side to a thermostat at zero temperature, inducing only friction, so that all rotors eventually come to rest. 
A lower bound for the dissipation rate, of a similar kind as \eqref{eq: effective dissipation h-m}, is derived mathematically 
and shown numerically to be optimal.  
 
If the chain \eqref{eq: usual chain} is connected at the two ends to reservoirs at different temperatures, 
it is expected to relax to a non-equilibrium stationary state (NESS). 
Moreover, for a fixed length, one would typically expect the relaxation to be exponentially fast.  
This is indeed what happens if $b < a$, 
i.e.\@ if the condition \eqref{eq: bad transport condition} is violated so that the pinning becomes irrelevant at high energies, 
see e.g.~\cite{cuneo_et_al_2018} and references therein. 
Instead, when \eqref{eq: bad transport condition} holds and when the convergence to the NESS can be established, 
it is shown in \cite{hairer_mattingly_2009} that the decay towards the NESS only follows a stretched exponential, due to the presence of breathers in the system.
The same conclusion holds for chains of three and four rotors \cite{cuneo_eckmann_poquet_2015,cuneo_eckmann_2016,cuneo_poquet_2017}
(the case of more than four rotors has not been investigated mathematically but a stretched exponential convergence is expected there as well).

\section{Quenched disordered systems}\label{sec: disordered systems}
In the rotor chain discussed in Section~\ref{sec: rotor chain}, the frequency mismatch between nearby rotors was due to thermal fluctuations. 
Another way to realize the same physical effect is to introduce quenched disorder, as e.g.\@ in the classical Hamiltonian 
\begin{equation}\label{eq: disordered chain}
	H(p,q) \; = \;  \sum_{x=1}^L \left( \frac{p_x^2}{2 m} + \frac{m\omega_x^2 q_x^2}{2} + g_0(q_x - q_{x+1})^2 + g \frac{q_x^4}{4} \right)
\end{equation}
with free boundary conditions ($q_{L+1} = q_L$) and 
where the frequencies $\omega_x$ of the on-site harmonic oscillators are taken to be strictly positive iid random variables with standard deviation $W > 0$.
For this system, it may seem that the frequency mismatch is much more robust than in the rotor chain \eqref{eq: rotors} 
since the frequencies $\omega_x$ do not evolve with time. 
This is indeed so in the purely harmonic case $g = 0$, where the Hamiltonian \eqref{eq: disordered chain} turns out to be an Anderson insulator in disguise. 
In particular, in $d=1$, all eigenmodes of the chain are localized for any value of $g_0$, 
implying a vanishing thermal conductivity \cite{bernardin_huveneers_2013} and a strong breakdown of local equilibrium \cite{de_roeck_dhar_huveneers_schutz_2017}, 
see also \cite{lepri_livi_politi_2003} and references theirein. 

However, the situation changes in the presence of anharmonic interactions, i.e.\@ $g > 0$, that provide effective couplings among the localized modes. 
It was observed numerically in \cite{dhar_lebowitz_2008} that 
``even a small amount of anharmonicity leads to a $J \sim 1/L$ dependence, implying diffusive transport of energy", 
see also \cite{dhar_saito_2008,mulanski_ahnert_pikovsky_shepelyansky_2009}.
Normal transport at any non-zero value of the coupling $g$ was confirmed in \cite{oganesyan_pal_huse_2009} for a classical spin chain,
where moreover a drastic reduction of the conductivity as $g \to 0$ is observed, 
``consistent with the possibility that the transport is actually essentially non-perturbative in $g$". 
This was justified mathematically in \cite{huveneers_nonlinearity_2013} in very much the same way as for the rotor chain discussed in Section~\ref{sec: rotor chain}: 
Fixing the disorder strength $W$ and taking here $g/\beta g_0^2$ instead of $g \beta$ as the perturbative parameter, 
the noise \eqref{eq: velocity flip} is added to the dynamics with a very slow rate \eqref{eq: strength noise}, 
leading to expressions analogous to \eqref{eq: math conductivity} and \eqref{eq: non math conductivity} for the conductivity
(see \cite{aoki_lukkarinen_spohn_2006} for scaling relations of the thermal conductivity with the various parameters of the model in the chain \eqref{eq: disordered chain}). 

However, the origin of transport in the chain \eqref{eq: disordered chain} is harder to elucidate than in the rotor chain. 
Indeed, there are now two types of chaotic spots: 
spots induced by fluctuations of the quenched disorder or by thermal fluctuations as in the rotor chain. 
The first one are abundant but they are obviously unable to travel through the chain. 
The second one can travel but require much more atypical thermal fluctuations than in the rotor chain to be formed, due to the presence of random frequencies. 
The detailed analysis carried in \cite{basko_2011} for the related DDNLS chain shows that the rate of dissipation is still of a strength comparable to \eqref{eq: basko rate}, 
i.e.\@ smaller than any power law, but larger than any exponential. 
More important, it turns out that here as well, thermal fluctuations furnish the fastest channel for transport. 
If this theory is correct, 
we are thus led to the conclusion that the presence of quenched disorder is eventually irrelevant 
and that the dynamical behavior becomes glassy in the chain \eqref{eq: disordered chain} as well, 
in very much the same sense as for the translation invariant systems. 

Finally, we would like to mention some puzzling observations on the spreading of an initially localized wave packet 
in a disordered medium where all other oscillators are initially at rest, 
as they might hint at the fact that some important aspect of the problem has been overlooked in the discussion so far. 
As is it the case for the chain in \eqref{eq: disordered chain},
we consider systems where the limit $g \to 0$ may be equivalently realized by taking the limit $\beta \to \infty$, 
which is relevant here since, if any local equilibrium sets-in, the local temperature must go to $0$. 
This set-up has been studied intensively. 
Theoretical \cite{fishman_krivolapov_soffer_2008,fishman_krivolapov_soffer_2009,fishman_krivolapov_soffer_2012}
and mathematical \cite{froehlich_spencer_wayne_1986,benettin_froehlich_giorgilli_1988,poschel_1990,wang_zhang_2009,bourgain_wang_2007} studies
all yield results on the slow, essentially non-perturbative, spreading of the wave packet as a function of time, 
while the evolution is fit with a power law in most numerical studies, 
see e.g.~\cite{pikovsky_shepelyansky_2008,flach_krimer_skokos_2009,ivanchenko_laptyeva_flach_2011,laptyeva_ivancheko_flach_2014,
mulansky_pikovsky_2010,mulansky_pikovsky_2013,vakulchyk_fistul_flach_2019}. 
Mathematical results are however hard to compare formally with numerical observations, 
since the former deal either with a transient regime or impose strong constraints on the parameters of the model. 
To see that these observations are conflicting with the arguments developed here, let us focus on the results in \cite{flach_krimer_skokos_2009,vakulchyk_fistul_flach_2019}, 
where the second moment $m_2$ of the wave packet is found to scale as $t^\alpha$ with $\alpha = 1/3$. 
Further, it is argued in \cite{flach_krimer_skokos_2009,vakulchyk_fistul_flach_2019} 
that this exponent describes the proper asymptotic long-time regime, that the value $\alpha = 1/3$ is universal, and that the
``spreading is due to corresponding weak chaos inside the packet, which slowly heats the cold exterior" \cite{flach_krimer_skokos_2009}. 
If the packet indeed spreads through local thermalization, 
then the observed scaling of $m_2$ is incompatible with the fact that the conductivity decays faster than any power law in the temperature.

\section{Quantum systems}\label{sec: quantum systems}
The main reason why quantum systems may behave differently from classical ones is that a few degrees of freedom cannot be properly chaotic, 
i.e.\@ the spectrum of a finite system remains discrete and cannot behave as a perfect bath. 
And indeed, as pointed out in the introduction, 
a specific class of systems was found to host an MBL phase, characterized by the total absence of transport of the conserved quantities and an emergent integrability, 
see e.g.\@ the reviews \cite{nandkishore_huse_2015,abanin_papic_2017,alet_florencie_2018}.
Systems known to display genuine MBL are one-dimensional quenched disordered lattice Hamiltonians with a finite Hilbert space per site, 
such as chains of spins or fermions on a lattice.
Extensions have naturally been investigated and in particular, the possibility of MBL in clean, translation invariant systems is of interest.  
Investigated cases include: 
systems with long range anharmonic interactions \cite{kagan_maksimov_1984} (even much before MBL was coined), 
long lived metastable states for lattice bosons \cite{carleo_becca_schiro_fabrizio_2012}, 
two species models \cite{grover_fisher_2014,shiulaz_mueller_2014,schiulaz_silva_mueller_2015}, 
Bose-Hubbard Hamiltonian at high density of particles \cite{de_roeck_huveneers_2014,bols_de_roeck_2018},
multi-lane models \cite{yao_et_al_2016},
quantized classical glasses \cite{hickey_genway_garrahan_2016}, 
and spin chains with strongly anharmonic interactions \cite{michailidis_et_al_2018}. 
As it turns out, these have only be found to exhibit slow dynamics but no true MBL phase
\cite{de_roeck_huveneers_prb_2014,de_roeck_huveneers_pspde_2015,papic_stoudenmire_abanin_2015}.  

To get one concrete example, let us consider the Bose-Hubbard Hamiltonian  
\begin{equation}\label{eq: bose hubbard hamiltonian}
	H \; = \; U \sum_{x=1}^L n_x^2 + g \sum_{x=1}^{L-1} (a_x^\dagger a_{x+1} + a_{x+1}^\dagger a_x ) 
\end{equation}
with $n_x = a_x^\dagger a_x$, that can be regarded as the exact quantum analog of the DNLS chain in \eqref{eq: dnls}.
Results on the non-perturbative character of the conductivity in the limit where the density of particles $\rho \to \infty$ have been derived in \cite{bols_de_roeck_2018}, 
improving on the techniques first developed in \cite{de_roeck_huveneers_2014}.
For the quantum chain \eqref{eq: bose hubbard hamiltonian}, 
there is no obvious way to introduce noise in the system 
and instead, to obtain a mathematical statement, the time-integral defining the conductivity \eqref{eq: Green Kubo} is truncated to a high power in $\rho$. 
The mathematical result in \cite{bols_de_roeck_2018} supports the claim that 
$$
	\rho^{r} \kappa (\rho) \to 0 \qquad \text{as} \qquad \rho \to \infty \qquad \text{for all} \qquad r \ge 0, 
$$ 
for fixed values of the temperature and the parameters of the model $g,U$.   

Let us next analyze why this system is not expected to be genuinely localized
(in particular why the conductivity does not vanish exactly starting from some density $\rho$).
Obviously, one may first argue that this is just impossible since the Hamiltonian in \eqref{eq: bose hubbard hamiltonian}
approaches its classical limit \eqref{eq: dnls} as $\rho \to \infty$.
This is true but this difficulty is superficial and can be circumvented.
For example, in the spirit of \cite{de_roeck_huveneers_prb_2014},
one may impose a cut-off on the total number of bosons per site, and consider the system in the regime $g/U \ll 1$ rather than in the limit $\rho \gg 1$. 
A similar set-up is considered in \cite{shiulaz_mueller_2014,schiulaz_silva_mueller_2015,michailidis_et_al_2018}.

This said, following \cite{de_roeck_huveneers_prb_2014,de_roeck_huveneers_pspde_2015}, 
let us now come to a reason for delocalization that applies to all models quoted above, 
but let us stick to the Hamiltonian \eqref{eq: bose hubbard hamiltonian} for commodity. 
Due to thermal fluctuations, one encounters regions where the density of particles is anomalously low, i.e.\@ much lower than $\rho$. 
There, the system is expected to be thermal and these spots act thus as (imperfect) baths. 
Even though they are imperfect, it should be realized that the level spacing decreases exponentially as the size of the rare regions grows, 
hence a large system contains unavoidably spots that can be considered effectively as good baths, 
in the sense that the sites in their vicinity can be thermalized. 
Therefore, by heating up near spins, the spot can migrate into the chain and modify the energy landscape along the way. 
We come thus to the conclusion that the main mechanism for energy transfer in strongly anharmonic quantum chains near the integrable limit
is properly the same as for classical systems, even though more space than two or three sites may be needed to realize a good enough effective bath. 

Finally, it is interesting to notice that rare regions may also have deep destabilizing effects on quenched disorder spin chains. 
First, the possibility of hot mobile bubbles, as described above, 
was shown to imply the absence of many-body mobility edges \cite{de_roeck_huveneers_muller_schiulaz_2016}, 
i.e.\@ that the presence of thermal states at a given value of the thermodynamical parameters implies that states are thermal at all values of these parameters. 
Second, while for classical systems we concluded in Section~\ref{sec: disordered systems} that the presence of chaotic spots due to fluctuations of the quenched disorder
did not play a major role in the transport of the energy, their presence become relevant for quantum systems: 
They destabilize the MBL phase in $d > 1$ \cite{de_roeck_huveneers_2017,luitz_huveneers_de_roeck_2017} 
and are expected to drive the system to the thermal phase if the bare localization length exceeds some threshold value in $d=1$ 
\cite{luitz_huveneers_de_roeck_2017,thiery_huveneers_muller_de_roeck_2018}.

% acknowledgements
\bigskip
\noindent
\textbf{Acknowledgements.}
This work of F.H. was partially supported by the grants ANR-15-CE40-0020-01 LSD and ANR-14-CE25-0011 EDNHS of the French National Research Agency (ANR). W.D.R. acknowledges the
support of the Flemish Research Fund FWO under grant
G076216N.

% bibliography
\bibliographystyle{plain}
\bibliography{kinetically_constrained_transport.bib}

\end{document}